\documentclass[preprint]{aastex}
\usepackage{amssymb,amsmath}
\usepackage{graphicx}
\usepackage{natbib}
\usepackage[OT1,T1]{fontenc}

\begin{document}

\title{RADYN simulations of non-thermal and thermal models of Ellerman bombs}

\author{Jie Hong\altaffilmark{1,2,3}, Mats Carlsson\altaffilmark{2}, and M.~D. Ding\altaffilmark{1,3}}
\affil{\altaffilmark{1}School of Astronomy and Space Science, Nanjing University, Nanjing 210023, China \email{dmd@nju.edu.cn}}
\affil{\altaffilmark{2}Institute of Theoretical Astrophysics, University of Oslo, P.O. Box 1029 Blindern, NO-0315 Oslo, Norway}
\affil{\altaffilmark{3}Key Laboratory for Modern Astronomy and Astrophysics (Nanjing University), Ministry of Education, Nanjing 210023, China}

\begin{abstract}
Ellerman bombs (EBs) are brightenings in the H$\alpha$ line wings that are believed to be caused by magnetic reconnection in the lower atmosphere. To study the response and evolution of the chromospheric line profiles, we perform radiative hydrodynamic simulations of EBs using both non-thermal and thermal models. Overall, these models can generate line profiles that are similar to observations. However, in non-thermal models we find dimming in the H$\alpha$ line wings and continuum when the heating begins, while for the thermal models dimming occurs only in the H$\alpha$ line core, and with a longer lifetime. This difference in line profiles can be used to determine whether an EB is dominated by non-thermal heating or thermal heating. In our simulations, if a higher heating rate is applied, the H$\alpha$ line will be unrealistically strong, while there are still no clear UV burst signatures.
\end{abstract}

\keywords{line: profiles --- radiative transfer --- Sun: activity --- Sun: atmosphere}

\section{Introduction}
Ellerman bombs (EBs) were first seen as transient brightenings in the H$\alpha$ and other Balmer lines \citep{1917ellerman}. The line wings of H$\alpha$ are enhanced, while the line center remains undisturbed. Observations have shown that EBs are also visible in other spectral lines and continua, including the \ion{Ca}{2} 8542 \AA\ line \citep{2013yang,2015kim,2015rezaei}, \ion{Ca}{2} H and K lines \citep{2008matsumoto,2015rezaei}, G band \citep{2011herlender,2013nelson}, ultraviolet continuum at 1600 and 1700 \AA\ \citep{2013vissers,2015rezaei,2016tian}, \ion{Mg}{2} triplet lines \citep{2015vissers,2017hansteen,2017hong}, and \ion{He}{1} D$_{3}$ and 10830 \AA\ lines \citep{2017libbrecht}. The typical lifetime of EBs is a few minutes and the typical size is about 1--2$\arcsec$ in diameter \citep{2002georgoulis,2013rutten}.

EBs are often observed near polarity inversion lines, with magnetic cancellation \citep{2008matsumotob,2013nelson}. The cancellation rate is estimated to be $10^{14}-10^{15}$ Mx s$^{-1}$ \citep{2016reid}. In observations, EBs are always associated with other activities including moat flows \citep{2011watanabe,2013vissers}, newly emerged magnetic flux \citep{2007pariat,2008watanabe,2013yang,2016reid,2016yang,2017danilovic}, jets \citep{2011watanabe,2015reid} or surges \citep{2008matsumoto,2013yang,2016pasechnik}. Numerical simulations show that magnetic reconnection of a U-shaped or $\Omega$-shaped loop in the lower atmosphere can be the cause of an EB \citep{2007isobe,2009archontis,2017hansteen}. Different methods to invert the spectral lines including H$\alpha$, \ion{Ca}{2} 8542 \AA, and \ion{Fe}{1} lines have been used to obtain the local temperature enhancement in the lower atmosphere. Both semi-empirical and two-cloud models show that the temperature enhancement is 600--3000 K \citep{2006fang,2006socas,2013gonzalez,2014berlicki,2015li,2016kondrashova,2014hong,2017hong}. 

Recently, the observation of UV bursts and the possible relation between EBs and UV bursts is a very hot topic. The main feature of UV bursts is the strong and wide line profile of \ion{Si}{4}, with a formation temperature of $8\times10^{4}$ K \citep{2014peter}. Previous observations show that EBs and UV bursts are related, suggesting that the temperature of EBs should be higher than the results from previous models \citep{2015kim,2015vissers,2016tian}. Recent numerical simulations \citep{2016ni} and spectral inversions of the \ion{He}{1} D$_{3}$ line \citep{2017libbrecht} show that it is possible to reach high temperatures in the lower atmosphere. However, using a semi-empirical model, \cite{2017fang} argued that the EB temperatures cannot be higher than 10000 K, in order to reproduce the observed line profiles and continuum. On the other hand, recent 3D simulations suggest that most UV bursts may originate from low/mid-chromospheric plasma \citep{2017hansteen}.

Up to now, there are very few radiative hydrodynamic simulations of EBs. \cite{2017reid} first calculated the line profiles of EBs using a thermal model, with results similar to observations. However, as EBs are believed to be caused by magnetic reconnection, the non-thermal effects by electrons, which they did not take into consideration, might be important. 

In this paper, we perform radiative hydrodynamic simulations of EBs using both non-thermal and thermal models, and make a comparison between these models, as well as between simulations and observations. We briefly introduce the simulation method in Section~2. In Section~3 we present the results of both non-thermal and thermal models. Then we compare the results and with observations in Section~4, and a summary is followed in Section~5.

\section{RADYN Simulations}
\subsection{Code Description}
The radiative hydrodynamics code RADYN was first developed by \cite{1992carlsson,1995carlsson,1997carlsson,2002carlsson} to study shocks in the chromosphere. More recently, it has been used to calculate the chromospheric response of a flare \citep{1999abbett,2005allred,2006allred,2015allred,2015arubio,2015brubio,2016rubio}. RADYN uses an adaptive grid \citep{1987dorfi} to solve the radiative hydrodynamics equations. Atoms that are important in the chromosphere are treated in non-LTE, including a hydrogen atom with six levels plus continuum, a singly ionized calcium atom with six levels plus continuum, and a helium atom with nine levels plus continuum. Complete frequency redistribution (CRD) is considered for all lines except the Lyman series, where the profiles are truncated at 10 Doppler widths to mimic the effect of partial frequency redistribution. The magnesium lines are not treated in detail here, because doing both CRD for calcium and magnesium lines will overestimate the net radiative rates \citep{2002uitenbroek}. All other atoms are included in the background opacity. A brief introduction of the flare version of the code can be found in \cite{2015allred}.

\subsection{Simulation Setup}
We assume a plane-parallel atmosphere with a quarter-circular loop structure. The loop length is set to 10 Mm from below the photosphere to the corona. The geometric structure of the loop is considered while calculating the X-ray ionizations and the gravitational acceleration of the upper atmosphere, which should have very little influence on the lower atmosphere which we are interested in.

\begin{figure}
\epsscale{0.8}
\plotone{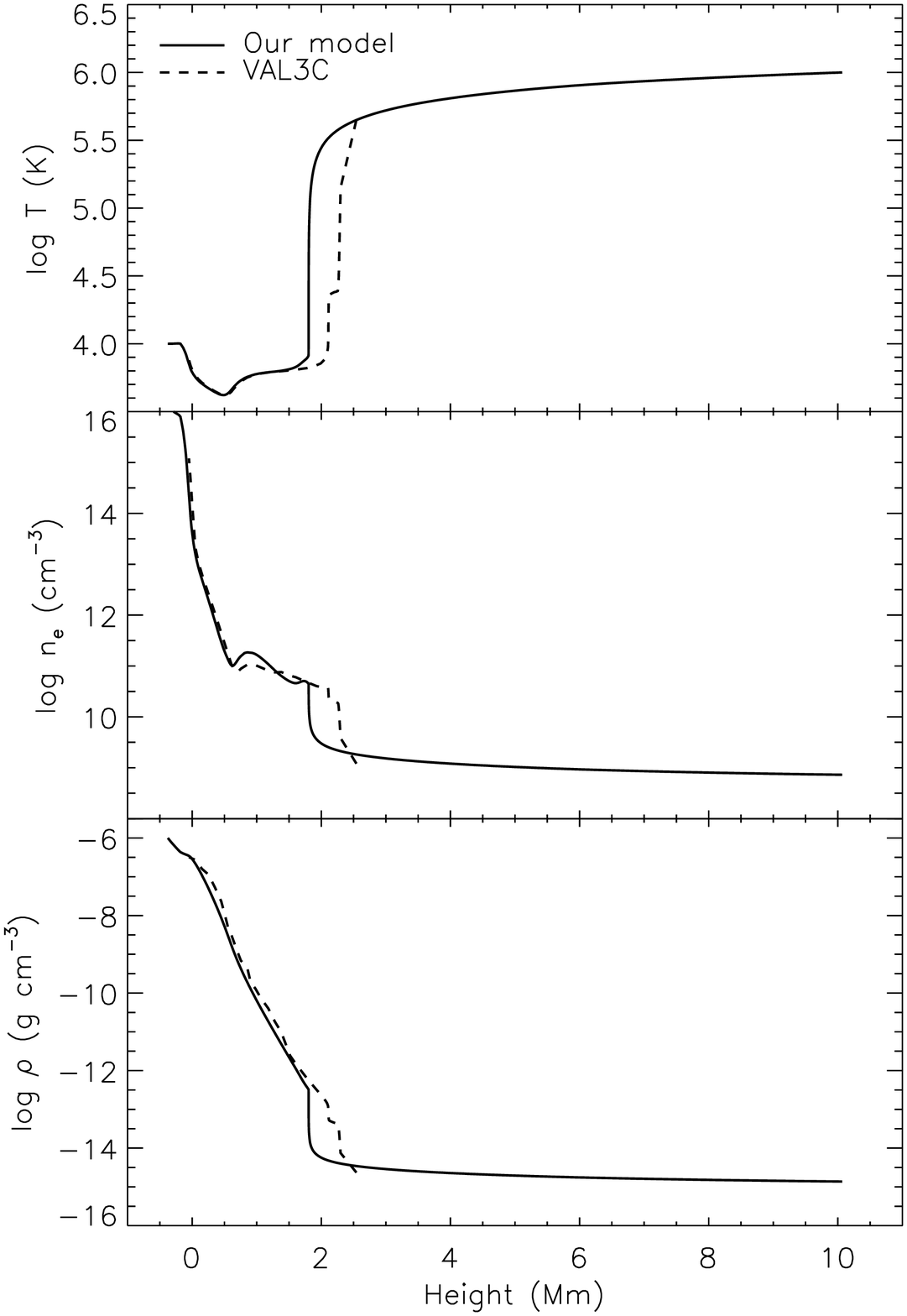}
\caption{Initial atmosphere of the RADYN simulations after relaxation. Height distributions of temperature (top), electron density (middle) and mass density (bottom) are shown as solid curves for our model and dashed curves for the VAL3C model, respectively.}\label{atm}
\end{figure}

The initial atmosphere for the simulations is generated based on the VAL3C model \citep{1981vernazza} but extended to 10 Mm. We add an extra energy term in the energy equation for the lower part of the atmosphere to balance the conductive and radiative losses, and we fix the temperature at the upper boundary. The whole atmosphere is then allowed to relax until near equilibrium is reached. The 1D atmosphere is discretized into 300 grid points. In Figure~\ref{atm}, we show the structure of temperature, electron density and mass density of the initial atmosphere and the VAL3C model.

\begin{figure}
\plotone{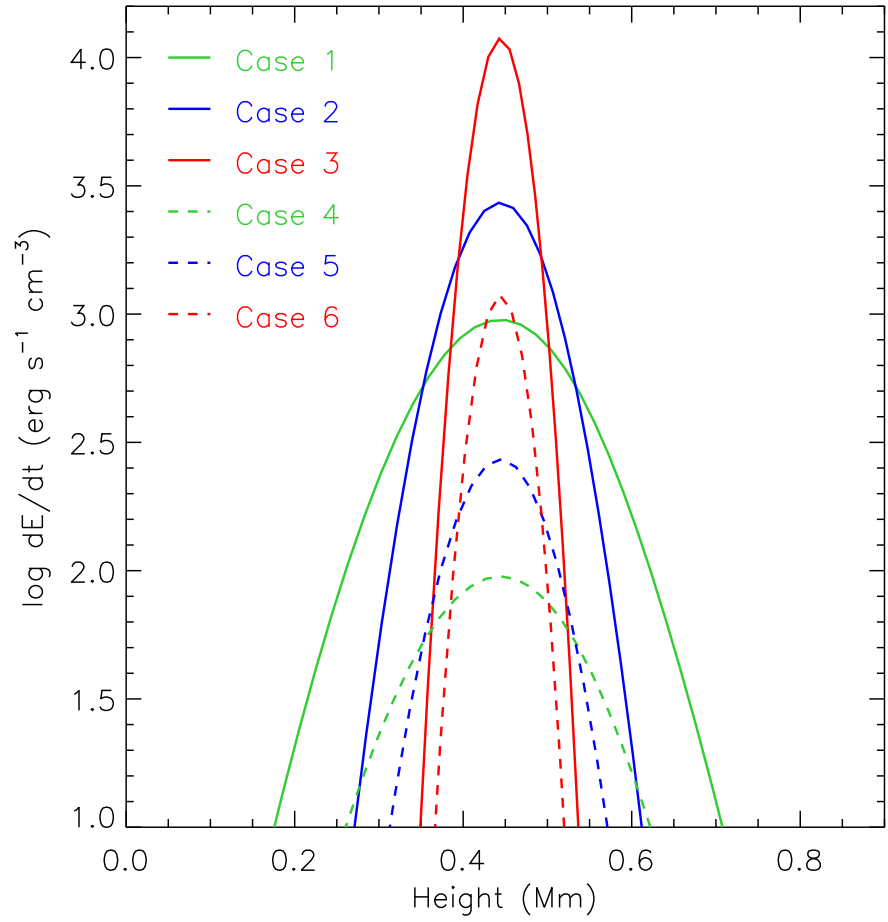}
\caption{Height distribution of beam heating rates at the beginning of the non-thermal simulations for the six cases.}\label{bhr}
\end{figure}

EBs are believed to be generated by magnetic reconnection in the lower atmosphere, where electrons are possibly accelerated to high energies. These non-thermal electrons can heat the plasma by Coulomb collisions and also cause non-thermal excitation and ionization of hydrogen atoms. This kind of process is mimicked by adding a heating rate from an electron beam in the energy equation and adding the non-thermal excitation and ionization rates in the rate equations, following \cite{1993fang}. The beam heating function is considered to have a Gaussian shape as a function of column mass, centered on the temperature minimum region (TMR), which is at 450 km in height and 0.1 g cm$^{-2}$ in column mass. We run six different non-thermal cases with varying shapes of the beam heating function in the lower atmosphere. The beam heating functions of the six cases are shown in Figure~\ref{bhr}. In Cases 4--6, the Gaussian shape has the same centering and width as in Cases 1--3, but the energy flux is one order less. By assuming a typical spatial size of EBs to be $1\arcsec.8\times1\arcsec.8$, we estimate that the energy rate of Case 6 is $1.6\times10^{22}$ erg s$^{-1}$, which is similar to the NICOLE inversion results ($2.2\times10^{22}$ erg s$^{-1}$; \citealt{2016reid}). For the thermal cases, we introduce a thermal heating rate in the energy equation and set the value to be the same as the beam heating rate in the non-thermal cases in order to make comparisons. Therefore, the difference between non-thermal and thermal models is that the non-thermal models include an extra excitation and ionization rate from the non-thermal electrons which influences the population densities. To distinguish between them, non-thermal cases are followed by the letter `a' while thermal cases are followed by the letter `b' in this paper.

We run all the twelve simulation cases with a closed upper boundary for 10 s. All non-thermal and thermal heating rates are time-independent, and last for the whole simulation time. We save the simulation snapshots every 0.1 s, and calculate the H$\alpha$ and \ion{Ca}{2} 8542 \r{A} line profiles of every snapshot.

\section{Results}

\subsection{Non-thermal Models}

\subsubsection{Dimming in H$\alpha$ and Continuum at 5000 \r{A}}
\begin{figure}
\plotone{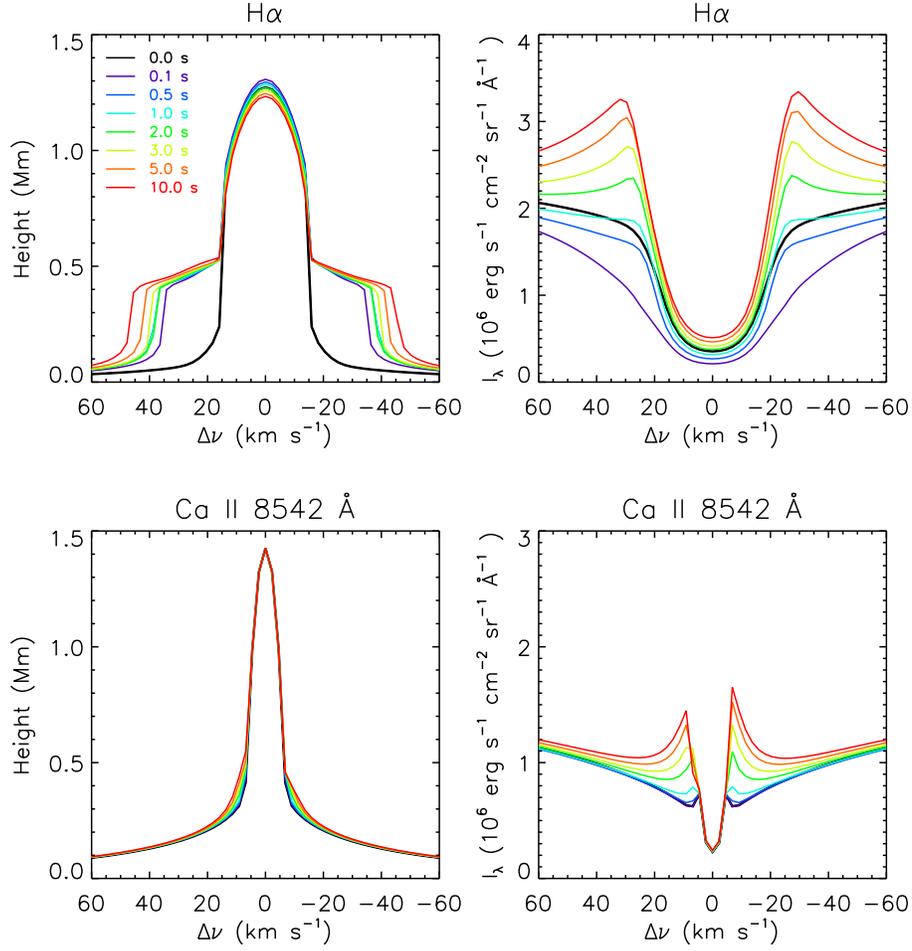}
\caption{Time evolution of the $\tau_{\nu}=1$ curve and the line profiles of H$\alpha$ and \ion{Ca}{2} 8542 \AA\ for Case 6a.}\label{tau6a}
\end{figure}

\begin{figure}
\plotone{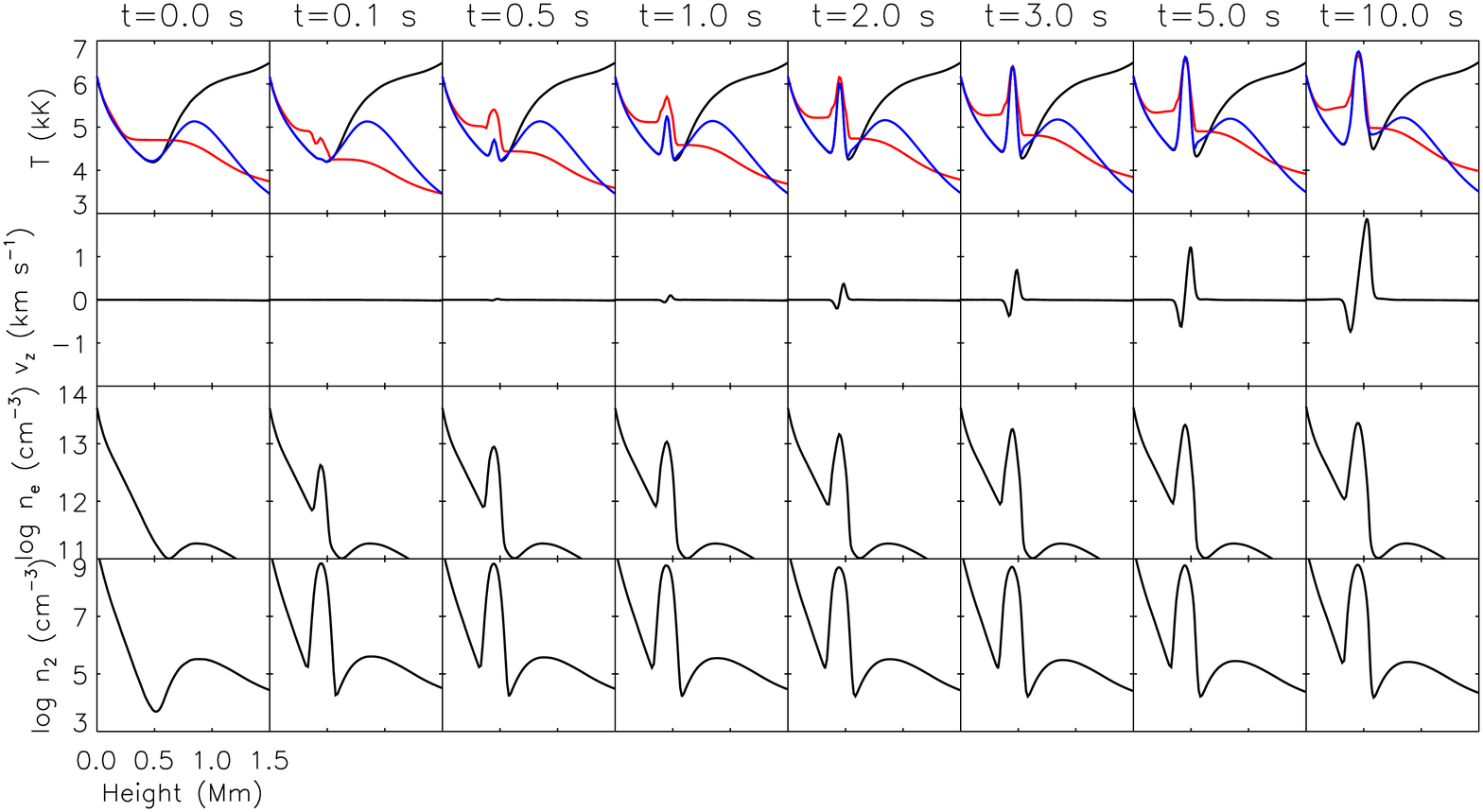}
\caption{Time evolution of the Planck function and line source functions, the vertical velocity, the electron density, and the hydrogen population density at the n=2 level. In the top row, the Planck function (black) and the line source functions (red for H$\alpha$, blue for \ion{Ca}{2} 8542 \r{A}) are plotted as an ``equivalent'' temperature, which is derived by equaling the source function to a Planck function at the same wavelength. In the second row, a positive velocity means an upflow, while a negative one means a downflow.}\label{atm6a}
\end{figure}

The most striking feature of the non-thermal model is the dimming in the H$\alpha$ line. We show the evolution of the $\tau=1$ height and the line profiles of Case 6a in Figure~\ref{tau6a}. The intensity of the H$\alpha$ line wings suddenly decreases when non-thermal heating begins. The decrease in H$\alpha$ intensity can be as large as 40\% at the EB peak wavelength ($-0.65$ \AA). At 0.1 s, there is a large amount of non-thermal electrons in the TMR (Figure~\ref{atm6a}), which collide with the neutral hydrogen atoms and can excite these atoms. Therefore, a rise in the population density of the n=2 level (the lower level of the H$\alpha$ line) causes the H$\alpha$ opacity to increase suddenly, giving the upward shift of the $\tau_{\nu}=1$ curve (Figure~\ref{tau6a}). The formation height of the line wings suddenly moves from the lower photosphere to the TMR, where the local source function is still very small, since the plasma has not been heated yet. Thus the emergent intensity suffers a decrease at first. The line center is also dimmed because the line source function at the formation height is also decreasing (Figure~\ref{atm6a}). However, when heating continues, the local source function at TMR starts to rise. Then, the line wing intensity increases slowly, and in the end, it becomes stronger than the original value.

\begin{figure}
\epsscale{0.8}
\plotone{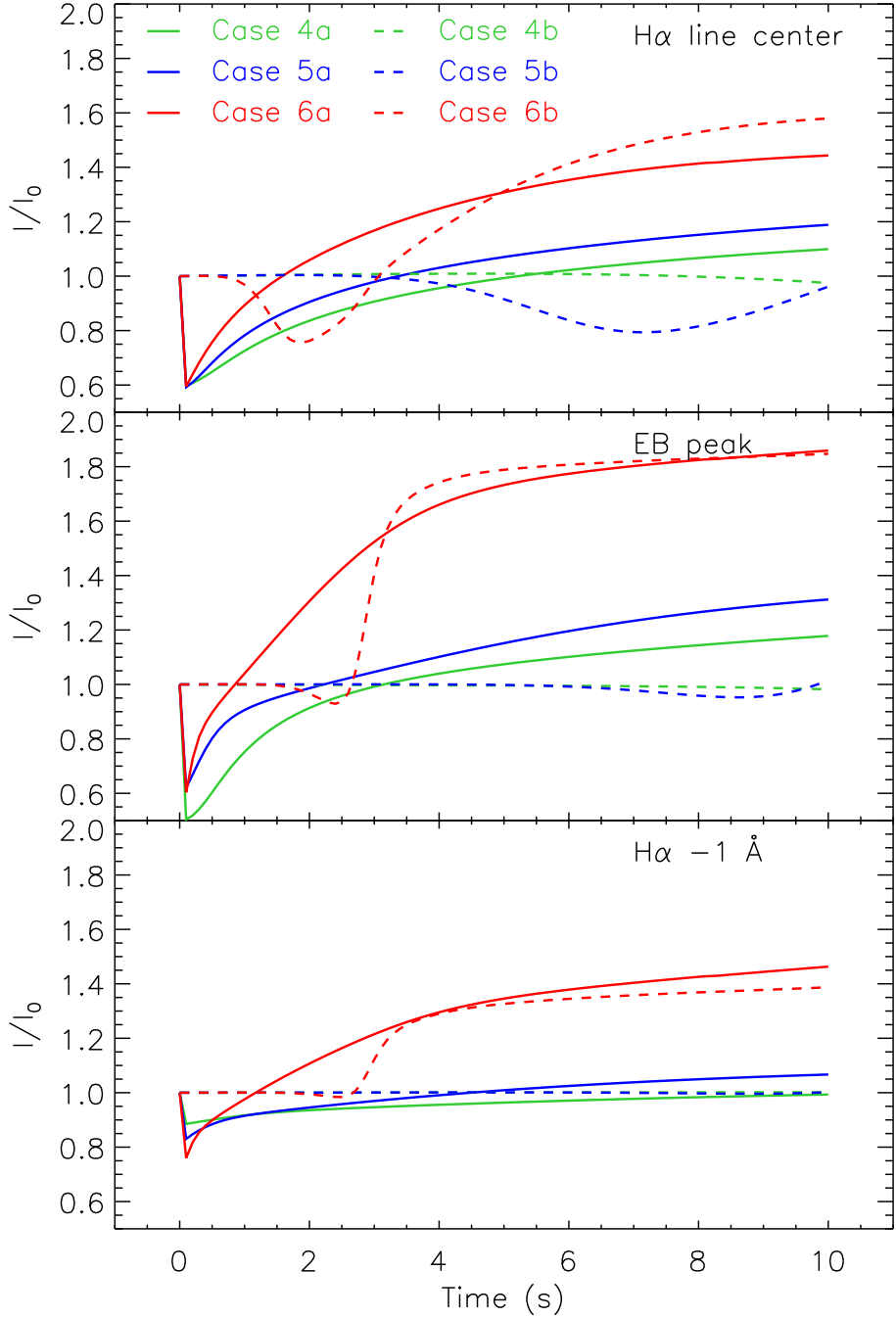}
\caption{Time evolution of the intensities at the H$\alpha$ line center, EB peak wavelength and line wing ($-1$ \AA), relative to the initial values. The EB peak wavelength is $-0.45$ \AA\ for Case 4, $-0.6$ \AA\ for Case 5, and $-0.65$ \AA\ for Case 6, respectively.}\label{evh}
\end{figure}

The dimming is seen in all the six non-thermal cases irrespective of how strong the beam heating is, and it starts just after the non-thermal heating begins. The beam heating rate determines how long the dimming lasts, which is dependent on how quick the line source function increases in the TMR. For the strongest Case 3a, the dimming ends after 0.2--0.3 s of heating; while for the weakest Case 4a, the dimming ends after about 6.5 s of heating (Figure~\ref{evh}). If we adopt a more gradual non-thermal heating rate like a triangular function over time instead of a rectangular function (constant within the heating period), the dimming effects still exist but are somewhat reduced.

At present, no systematic observations of EBs at the optical continuum are available for constraining models. Here, we also study the continuum at 5000 \AA, a wavelength representative of the visible waveband. This continuum is mostly of photospheric origin and comes from hydrogen recombination (Paschen continuum) and $H_{bf}^{-}$ emission. It is seen that the behavior of the continuum at 5000 \r{A} is similar to the H$\alpha$ line wings, but the decrease in intensity is only 2--3\% (Figure~\ref{evh}). As for the \ion{Ca}{2} 8542 \AA\ line, since it is more sensitive to the local temperature, we do not see any dimming in all cases presented here (Figure~\ref{evc}).

\subsubsection{EB Line Profiles}
As seen from the line profiles of Case 6a after 10.0 s of heating (Figure~\ref{tau6a}), there is a clear emission in the H$\alpha$ line wings, about 80\% increase in intensity at the EB peak wavelength ($-0.65$ \AA) compared to the original profile. The $\tau_{\nu}=1$ curve shows two bumps at the near wings, corresponding to the position where the line wing intensity is enhanced. This indicates that the formation height of the near wing is levitated because of the increase in the opacity near the TMR. This opacity increase is caused by the temperature increase and also the non-thermal excitation effect. It is seen that the temperature is increased by about 2500 K in the TMR (Figure~\ref{atm6a}). The high gas pressure in the TMR results in mass flows in both directions of about 2 km s$^{-1}$, which is also reflected in the weak line asymmetries. These bidirectional flows have also been observed and modeled previously \citep{2008watanabe,2008matsumoto,2009archontis,2017reid,2017hansteen}. One can also see that the line source function of the H$\alpha$ line decouples from the local Planck function for layers above 100 km, although the heating at the TMR reduces this decoupling.

Unlike that of the H$\alpha$ line, the line source function of the \ion{Ca}{2} 8542 \AA\ line couples with the Planck function in the TMR, indicating that the line wing intensity is mainly dependent on the local electron temperature. There is a very strong emission in the line wings (80\% increase in intensity), and the line profile shows a clear blue asymmetry.

We should also note that there is a large enhancement in the intensity of the H$\alpha$ line center in the modeling results. Such a discrepancy with observations is possibly due to the limitation of one-dimensional models since the three dimensional radiative effects may smooth out the emissivity at the line core formation height \citep{2012leenaarts}. The \ion{Ca}{2} 8542 \AA\ line is less influenced because the source function is strongly coupled to the temperature. Moreover, in our simulations, we do not assume an overlying fibrilar canopy in the upper chromosphere, which has been found in observations \citep{2011watanabe,2013rutten}. Such an undisturbed canopy can reduce the line center intensity of these two chromospheric lines.

\subsection{Thermal Models}

\subsubsection{Dimming in H$\alpha$ and EB line profiles}
\begin{figure}
\plotone{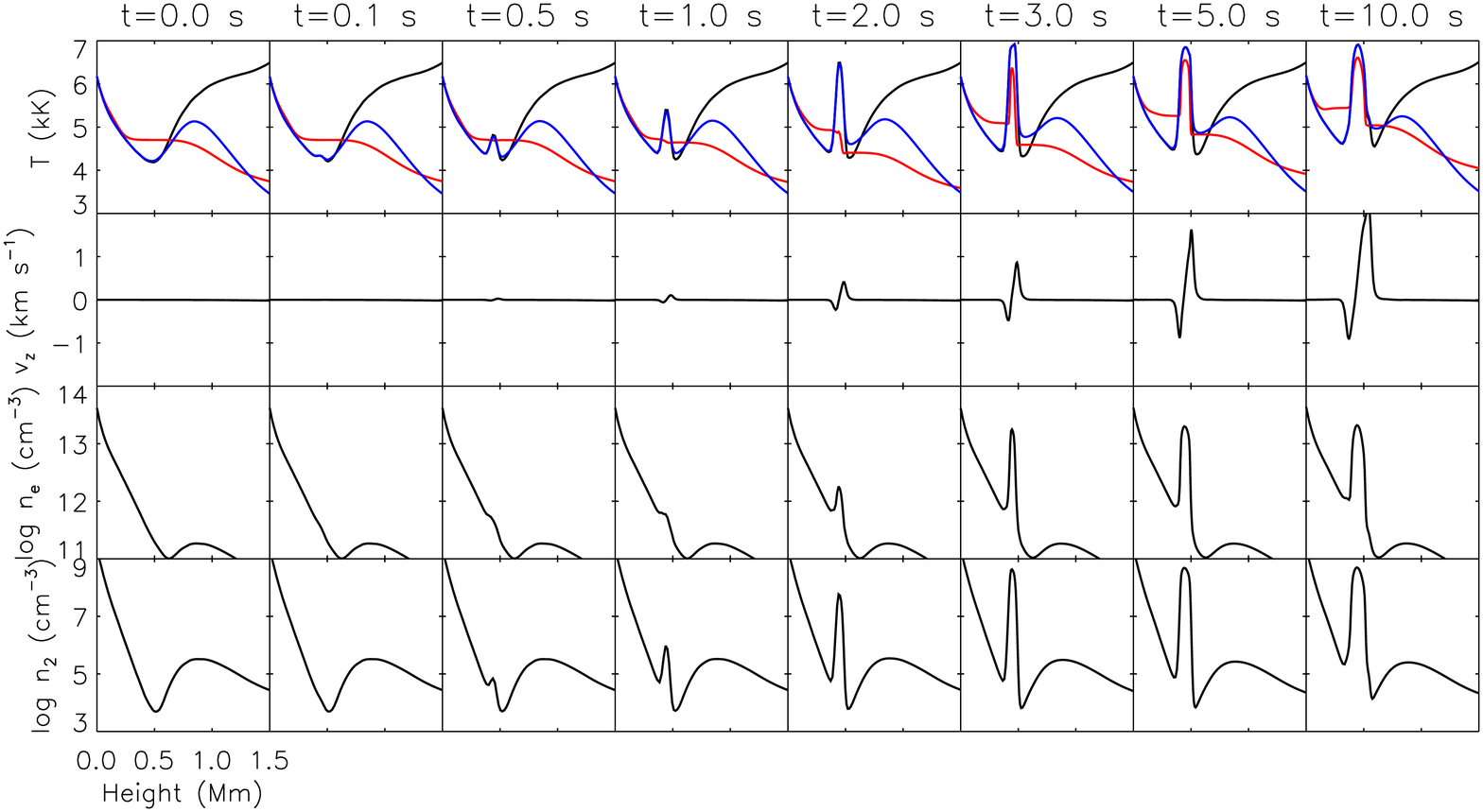}
\caption{Same as Fig.~\ref{atm6a}, but for Case 6b.}\label{atm6b}
\end{figure}

\begin{figure}
\plotone{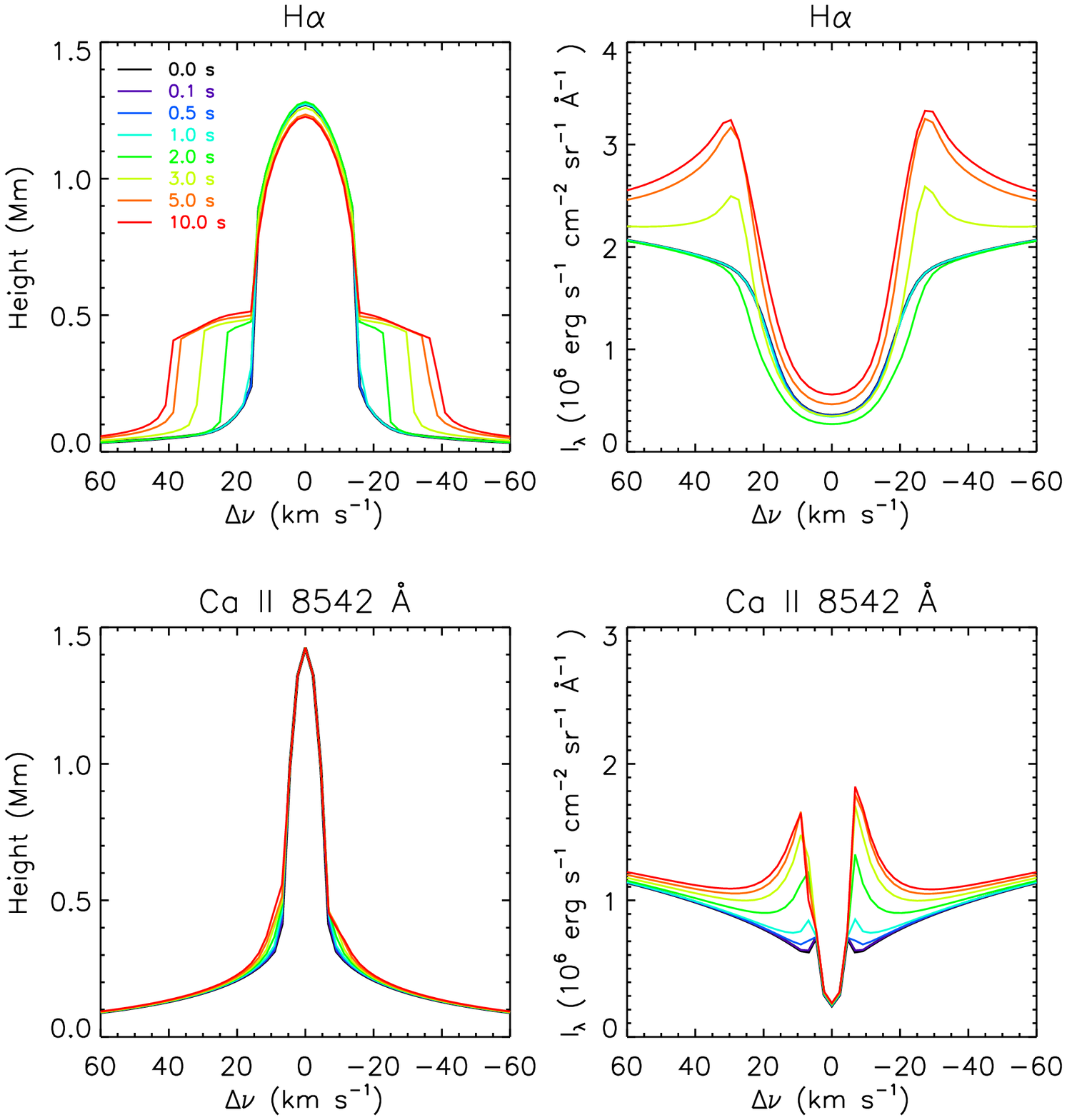}
\caption{Same as Fig.~\ref{tau6a}, but for Case 6b.}\label{tau6b}
\end{figure}

As for Case 6b, when thermal heating begins, there is not a significant increase in the local electron density (Figure~\ref{atm6b}), and the line profiles show no dimming (Figure~\ref{tau6b}). However, after 0.5 s of heating, a rise in the local temperature causes an increase in the hydrogen level population at n=2. The formation height of the near wings is thus levitated to the TMR and the line source function at the formation height of the line center is also decreasing, leading to a dimming most apparent at the line center and near wings, but not the far wings. The intensity at the line center and near wings reaches the minimum at 2.0 s, and begins to rise afterwards. The largest decrease is about 20\% at EB peak wavelength (Figure~\ref{evh}). This kind of dimming in H$\alpha$ is seen in all six thermal cases, and the heating rate determines when the dimming begins and how long it lasts. A larger thermal heating rate always means an earlier and shorter dimming in H$\alpha$. However, we do not see any dimming features in \ion{Ca}{2} 8542 \AA\ or the continuum at 5000 \r{A} (Figure~\ref{evc}).

After 10.0 s of heating, the temperature enhancement in the TMR is about 3000 K (Figure~\ref{atm6b}). The shapes of the $\tau_{\nu}=1$ curve and the line profiles are similar to the non-thermal cases. There is also a blue asymmetry caused by the mass flows.

\section{Discussion}
\subsection{Comparison between non-thermal and thermal models}
\begin{figure}
\epsscale{0.8}
\plotone{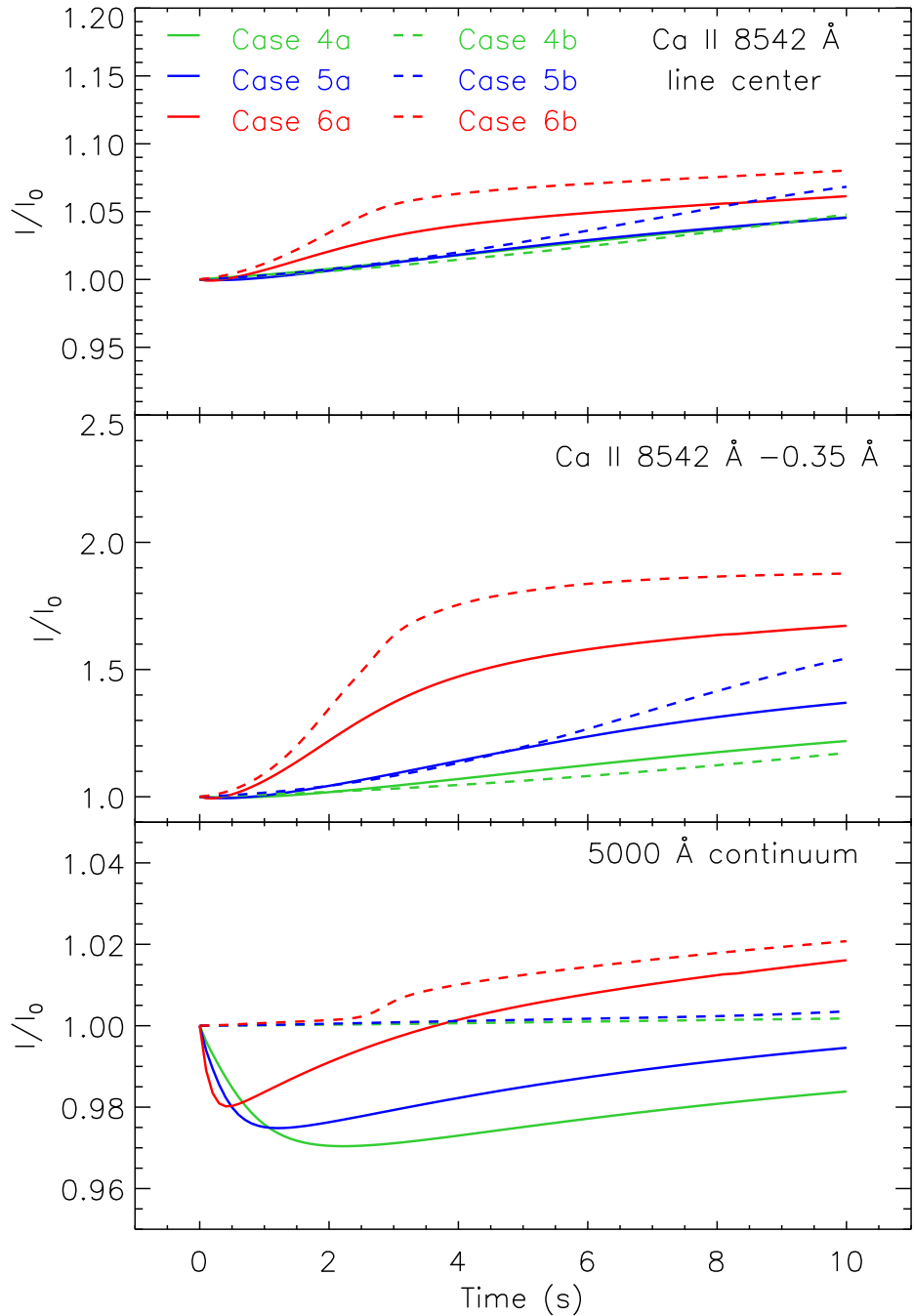}
\caption{Time evolution of the intensities at the \ion{Ca}{2} 8542 \r{A} line center and line wing ($-0.35$ \AA), and the 5000 \AA\ continuum, relative to the initial values.}\label{evc}
\end{figure}

The difference between the results from non-thermal and thermal models lies in the time evolution of the H$\alpha$ line profiles as well as continuum intensity (Figures~\ref{evh} and \ref{evc}). Although the H$\alpha$ dimming is seen in both cases, there are obvious differences in four aspects.

Firstly, in the non-thermal cases, the dimming begins instantly after the non-thermal heating, and the line intensity starts to rise afterwards; while in the thermal cases, the dimming begins much later in time. For example, in Case 6a, the line intensity decreases from the beginning suddenly, then increases gradually, and at 1.0 s reaches the original level. In Case 6b, however, the line intensity decreases gradually after 1.0 s and reaches its minimum at 1.8 s; then it goes back gradually, and at 2.6 s it recovers to the original level. This implies that the H$\alpha$ dimming lasts longer in the thermal case. That is the reason why after a heating for 10.0 s, the H$\alpha$ line intensity in Cases 4b and 5b is still increasing, while in Cases 4a and 5a, it seems to be saturated and does not change anymore. 
Secondly, dimming in the line center is much stronger in the non-thermal cases than that in the thermal cases. The relative decrease in line center intensity is 40\% for Case 6a, while it is only 25\% for Case 6b. Thirdly, there is also a slight dimming in the optical continuum in the non-thermal cases, with the largest intensity decrease being about 5\%. The dimming also appears in the far wings of H$\alpha$. However, as for the thermal cases, there is no dimming seen in the continuum and the H$\alpha$ far wings. 
Fourthly, the behavior of the local source function is different. In the non-thermal cases, the local source function rises first, before the rise of the Planck function (local temperature); while in the thermal cases, the local source function rises a little bit later than the Planck function (Figures~\ref{atm6a} and \ref{atm6b}). The reason is that the non-thermal heating instantaneously influences the local population density and also the line source function by non-thermal exitation and ionization rates, while the thermal heating mainly enhances the local temperature.

The direct cause for dimming in H$\alpha$ in both cases is an upward shift in the formation height of the near wings and a decrease of the line source function in the upper atmosphere (where the line center is formed). The near wings are formed higher in EBs relative to the quiet region because the level populations of hydrogen at excited levels, as well as the line opacity, are increased. In non-thermal cases, the inclusion of non-thermal rates increases the level populations of hydrogen (say, at n=2) effectively, which can cause a significant dimming in the line and the continuum. The electron beam induced dimming in the continuum has previously been revealed in solar flares \citep{1999abbett,2005allred,2006allred}. In thermal cases, the level populations of hydrogen at excited levels are increased mainly due to the increase of the local temperature. 
It is known that thermal excitation is less effective than non-thermal excitation. Moreover, in thermal cases, while the line opacity is increased, the line source function is also increased, which reduces the dimming effect. Therefore, the dimming in thermal cases only appears in the H$\alpha$ line core and is far less obvious than in the non-thermal cases. The continuum intensity in thermal models rises very slowly in the first 2 s (Figure~\ref{evc}), which is also a result of the mutual cancelation of the effects of an increased opacity and an increased source function at the continuum.

We also find that in the non-thermal cases, the continuum dimming lasts longer than the H$\alpha$ dimming. This is because the H$\alpha$ line source function has a weaker coupling to the local temperature than the continuum. As seen from Figure~\ref{atm6a}, the H$\alpha$ line source function rises before the rise of temperature, which means that the H$\alpha$ line can recover from dimming earlier than the continuum.

However, after a sufficient time of heating, the H$\alpha$ line profiles in the two models are nearly the same. Moreover, for the \ion{Ca}{2} 8542 line where the non-thermal effects do not play a major role, there appears to be very little difference between the two models in both the line profiles and their time evolution.

\subsection{Comparison between simulations and observations}
The H$\alpha$ and \ion{Ca}{2} 8542 \AA\ line profiles from our models show wing enhancements that are similar to observations. Case 4 displays the features of a weak EB with a temperature increase of 1000 K, while Case 6 displays a very strong EB with a temperature increase of about 3000 K.

An interesting feature present in the simulations is the dimming in H$\alpha$ line profiles in both non-thermal and thermal cases. It should be noted that a dimming in the H$\alpha$ line and also the continuum has been reported in non-thermal models of solar flares \citep{1991heinzel,1999abbett,2005allred,2006allred}. Clear dimming in the \ion{He}{1} 10830 \AA\ line was recently observed in a flare by \cite{2016xu}. However, in the previous EB observations, no dimming has been reported yet, possibly because the lifetime of the dimming is quite short. The dimming phase of non-thermal cases lasts less than a few seconds, which is comparable to or even shorter than the cadence of some instruments. Thus it is very difficult to catch the dimming. The slim chance calls for a large amount of observations, especially for weak EBs, because their dimming phase lasts longer. With observations at the line center, near wings, far wings of H$\alpha$ and even the continuum, we can determine whether the observed EB is dominated by non-thermal heating or thermal heating. We would like to point out that in some spectral observations, the cadence is relatively low if using a scanning technique over wavelength, where the dimming in H$\alpha$ might also contribute to the line asymmetries in addition to what is caused by the bidirectional mass flows. As for the dimming feature in the continuum, the intensity change is less than 5\% and thus could be buried in the background fluctuations of the granular intensity.

\begin{figure}
\plotone{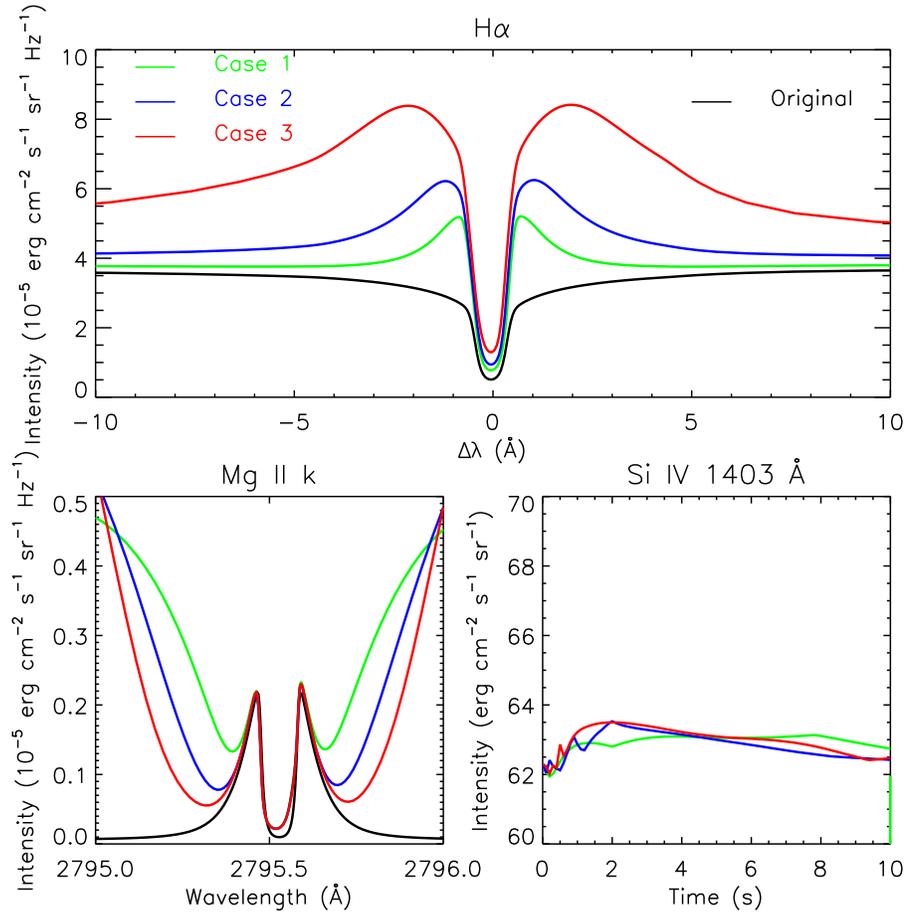}
\caption{Line profiles of H$\alpha$ (top) and \ion{Mg}{2} k (bottom left) at 10.0 s, as well as the evolution of the \ion{Si}{4} 1403 \AA\ line intensity integrated over wavelength (bottom right) for Cases 1--3.}\label{mg}
\end{figure}

Recently, many observations point to a possible relationship between EBs and UV bursts. It is therefore an interesting question if UV bursts also occur in the lower atmosphere, with a large local temperature increase in order to produce the UV emissions. We try to simulate some extreme cases with a very large beam heating rate near the TMR. Figure~\ref{mg} shows the H$\alpha$ and \ion{Mg}{2} k line profiles as well as the integrated intensity of the \ion{Si}{4} 1403 \AA\ line in Cases 1--3 after 10.0 s of heating. The \ion{Mg}{2} k line profiles are computed using the RH code \citep{2001uitenbroek,2015pereira}, and the integrated intensity of the \ion{Si}{4} 1403 \AA\ line over wavelength is calculated using CHIANTI \citep{2015delzanna}. One can see that similar to \cite{2017reid}, the outer wings of the \ion{Mg}{2} k line are enhanced, while there is very little influence on the k$_{2}$ peaks. The lightcurve of the \ion{Si}{4} integrated intensity shows very little fluctuation that is atypical in UV bursts. Besides, the H$\alpha$ line is unrealistically enhanced. It is seen that the line wing emission in Case 3 is extremely strong and stretches even beyond $\pm10.0$ \AA. The relative increase in intensity is more than 150\% at line center and more than 200\% at $\pm1.0$ \AA. In observations, an event with such features is of course far beyond the reasonable range of typical EBs.

\section{Summary}
We perform radiative hydrodynamic simulations of EBs using both non-thermal and thermal models. The energy conversion rate adopted in these models correspond to previous calculations and estimates for the energy budget of typical EBs. Basically, our simulations can generate typical EB wing enhancement of H$\alpha$ and \ion{Ca}{2} 8542 \AA\ comparable to observations. The discrepancy at the line center is possibly due to the assumption of one-dimensional models and the lack of a chromospheric canopy in the atmosphere. The temperature increase in the TMR is between 1000 and 3000 K, similar to previous results. If adopting a higher energy input rate, we can get a higher temperature increase in the TMR, which, however, produces an H$\alpha$ line unrealistically stronger than the observed one. The strongest cases here are all without UV burst signatures.

The most interesting point is that the models predict a dimming in the H$\alpha$ line at the beginning of the EBs. In non-thermal models, the dimming appears both in the H$\alpha$ line center and line wings when the heating begins. The line intensity can decrease by up to 50\%. The larger the beam heating rate is, the less time the dimming lasts. The period of dimming is usually less than a few seconds. The dimming also appears in the continuum, which is less strong but lasts longer than the H$\alpha$ dimming. However, no obvious dimming is found in the \ion{Ca}{2} 8542 \AA\ line.

There is also a dimming in the H$\alpha$ line in the thermal models, but it begins later and lasts longer than in the non-thermal models. Only the intensity at the line center and near wings decreases, with a magnitude less than in the non-thermal models. By comparison, there is no dimming in the continuum or the \ion{Ca}{2} 8542 \AA\ line.

In the future, it is possible to determine whether an EB is dominated by non-thermal heating or thermal heating by checking the H$\alpha$ line profile at the very beginning of EBs. For this purpose, we require observations of EBs with very high cadence and spatial resolution. On the other hand, we can also make a large amount of observations and search for possible dimming signatures in some EBs considering that the lifetime of dimming is quite short.

\acknowledgments
We thank the referee for constructive suggestions that helped improve the paper. This work was supported by NSFC under grants 11373023, 11403011, and 11533005, and NKBRSF under grant 2014CB744203 and by the Research Council of Norway. J.H. was also supported by CSC under file no. 201606190130. J.H. would like to thank Yuhao Zhou for helpful discussions.

\clearpage

\begin{thebibliography}{dummy}
\bibitem[Abbett \& Hawley(1999)]{1999abbett} Abbett, W.~P., \& Hawley, S.~L.\ 1999, \apj, 521, 906 
\bibitem[Allred et al.(2005)]{2005allred} Allred, J.~C., Hawley, S.~L., Abbett, W.~P., \& Carlsson, M.\ 2005, \apj, 630, 573 
\bibitem[Allred et al.(2006)]{2006allred} Allred, J.~C., Hawley, S.~L., Abbett, W.~P., \& Carlsson, M.\ 2006, \apj, 644, 484 
\bibitem[Allred et al.(2015)]{2015allred} Allred, J.~C., Kowalski, A.~F., \& Carlsson, M.\ 2015, \apj, 809, 104
\bibitem[Archontis \& Hood(2009)]{2009archontis} Archontis, V., \& Hood, A.~W.\ 2009, \aap, 508, 1469
\bibitem[Bello Gonz{\'a}lez et al.(2013)]{2013gonzalez} Bello Gonz{\'a}lez, N., Danilovic, S., \& Kneer, F.\ 2013, \aap, 557, A102
\bibitem[Berlicki \& Heinzel(2014)]{2014berlicki} Berlicki, A., \& Heinzel, P.\ 2014, \aap, 567, A110
\bibitem[Carlsson \& Stein(1992)]{1992carlsson} Carlsson, M., \& Stein, R.~F.\ 1992, \apjl, 397, L59 
\bibitem[Carlsson \& Stein(1995)]{1995carlsson} Carlsson, M., \& Stein, R.~F.\ 1995, \apjl, 440, L29 
\bibitem[Carlsson \& Stein(1997)]{1997carlsson} Carlsson, M., \& Stein, R.~F.\ 1997, \apj, 481, 500 
\bibitem[Carlsson \& Stein(2002)]{2002carlsson} Carlsson, M., \& Stein, R.~F.\ 2002, \apj, 572, 626 
\bibitem[Danilovic et al.(2017)]{2017danilovic} Danilovic, S., Solanki, S.~K., Barthol, P., et al.\ 2017, \apjs, 229, 5 
\bibitem[Del Zanna et al.(2015)]{2015delzanna} Del Zanna, G., Dere, K.~P., Young, P.~R., Landi, E., \& Mason, H.~E.\ 2015, \aap, 582, A56 
\bibitem[Dorfi \& Drury(1987)]{1987dorfi} Dorfi, E.~A., \& Drury, L.~O.\ 1987, Journal of Computational Physics, 69, 175 
\bibitem[Ellerman(1917)]{1917ellerman} Ellerman, F.\ 1917, \apj, 46, 298
\bibitem[Fang et al.(1993)]{1993fang} Fang, C., Henoux, J.~C., \& Gan, W.~Q.\ 1993, \aap, 274, 917 
\bibitem[Fang et al.(2006)]{2006fang} Fang, C., Tang, Y.~H., Xu, Z., Ding, M.~D., \& Chen, P.~F.\ 2006, \apj, 643, 1325
\bibitem[Fang et al.(2017)]{2017fang} Fang, C., Hao, Q., Ding, M.~D., \& Li, Z.\ 2017, arXiv:1702.01905 
\bibitem[Georgoulis et al.(2002)]{2002georgoulis} Georgoulis, M.~K., Rust, D.~M., Bernasconi, P.~N., \& Schmieder, B.\ 2002, \apj, 575, 506
\bibitem[Hansteen et al.(2017)]{2017hansteen} Hansteen, V.~H., Archontis, V., Pereira, T.~M.~D., et al.\ 2017, \apj, 839, 22 
\bibitem[Heinzel(1991)]{1991heinzel} Heinzel, P.\ 1991, \solphys, 135, 65 
\bibitem[Herlender \& Berlicki(2011)]{2011herlender} Herlender, M., \& Berlicki, A.\ 2011, Central European Astrophysical Bulletin, 35, 181 
\bibitem[Hong et al.(2014)]{2014hong} Hong, J., Ding, M.~D., Li, Y., Fang, C., \& Cao, W.\ 2014, \apj, 792, 13
\bibitem[Hong et al.(2017)]{2017hong} Hong, J., Ding, M.~D., \& Cao, W.\ 2017, \apj, 838, 101 
\bibitem[Isobe et al.(2007)]{2007isobe} Isobe, H., Tripathi, D., \& Archontis, V.\ 2007, \apjl, 657, L53
\bibitem[Kim et al.(2015)]{2015kim} Kim, Y.-H., Yurchyshyn, V., Bong, S.-C., et al.\ 2015, \apj, 810, 38
\bibitem[Kondrashova(2016)]{2016kondrashova} Kondrashova, N.~N.\ 2016, Kinematics and Physics of Celestial Bodies, 32, 70
\bibitem[Leenaarts et al.(2012)]{2012leenaarts} Leenaarts, J., Carlsson, M., \& Rouppe van der Voort, L.\ 2012, \apj, 749, 136 
\bibitem[Li et al.(2015)]{2015li} Li, Z., Fang, C., Guo, Y., et al.\ 2015, Research in Astronomy and Astrophysics, 15, 1513
\bibitem[Libbrecht et al.(2017)]{2017libbrecht} Libbrecht, T., Joshi, J., Rodr{\'{\i}}guez, J.~d.~l.~C., Leenaarts, J., \& Ramos, A.~A.\ 2017, \aap, 598, A33 
\bibitem[Matsumoto et al.(2008a)]{2008matsumoto} Matsumoto, T., Kitai, R., Shibata, K., et al.\ 2008, \pasj, 60, 95
\bibitem[Matsumoto et al.(2008b)]{2008matsumotob} Matsumoto, T., Kitai, R., Shibata, K., et al.\ 2008, \pasj, 60, 577 
\bibitem[Nelson et al.(2013)]{2013nelson} Nelson, C.~J., Shelyag, S., Mathioudakis, M., et al.\ 2013, \apj, 779, 125
\bibitem[Nelson et al.(2015)]{2015nelson} Nelson, C.~J., Scullion, E.~M., Doyle, J.~G., Freij, N., \& Erd{\'e}lyi, R.\ 2015, \apj, 798, 19
\bibitem[Ni et al.(2016)]{2016ni} Ni, L., Lin, J., Roussev, I.~I., \& Schmieder, B.\ 2016, \apj, 832, 195 
\bibitem[Pariat et al.(2007)]{2007pariat} Pariat, E., Schmieder, B., Berlicki, A., et al.\ 2007, \aap, 473, 279
\bibitem[Pasechnik(2016)]{2016pasechnik} Pasechnik, M.~N.\ 2016, Kinematics and Physics of Celestial Bodies, 32, 55
\bibitem[Pereira \& Uitenbroek(2015)]{2015pereira} Pereira, T.~M.~D., \& Uitenbroek, H.\ 2015, \aap, 574, A3 
\bibitem[Peter et al.(2014)]{2014peter} Peter, H., Tian, H., Curdt, W., et al.\ 2014, Science, 346, 1255726
\bibitem[Reid et al.(2015)]{2015reid} Reid, A., Mathioudakis, M., Scullion, E., et al.\ 2015, \apj, 805, 64
\bibitem[Reid et al.(2016)]{2016reid} Reid, A., Mathioudakis, M., Doyle, J.~G., et al.\ 2016, \apj, 823, 110
\bibitem[Reid et al.(2017)]{2017reid} Reid, A., Mathioudakis, M., Kowalski, A., Doyle, J.~G., \& Allred, J.~C.\ 2017, \apjl, 835, L37 
\bibitem[Rezaei \& Beck(2015)]{2015rezaei} Rezaei, R., \& Beck, C.\ 2015, \aap, 582, A104
\bibitem[Rubio da Costa et al.(2015a)]{2015arubio} Rubio da Costa, F., Kleint, L., Petrosian, V., Sainz Dalda, A., \& Liu, W.\ 2015, \apj, 804, 56 
\bibitem[Rubio da Costa et al.(2015b)]{2015brubio} Rubio da Costa, F., Liu, W., Petrosian, V., \& Carlsson, M.\ 2015, \apj, 813, 133 
\bibitem[Rubio da Costa et al.(2016)]{2016rubio} Rubio da Costa, F., Kleint, L., Petrosian, V., Liu, W., \& Allred, J.~C.\ 2016, \apj, 827, 38 
\bibitem[Rutten et al.(2013)]{2013rutten} Rutten, R.~J., Vissers, G.~J.~M., Rouppe van der Voort, L.~H.~M., S{\"u}tterlin,
P., \& Vitas, N.\ 2013, Journal of Physics Conference Series, 440, 012007
\bibitem[Rutten et al.(2015)]{2015rutten} Rutten, R.~J., Rouppe van der Voort, L.~H.~M., \& Vissers, G.~J.~M.\ 2015, \apj, 808, 133
\bibitem[Socas-Navarro et al.(2006)]{2006socas} Socas-Navarro, H., Mart{\'{\i}}nez Pillet, V., Elmore, D., et al.\ 2006, \solphys, 235, 75
\bibitem[Tian et al.(2016)]{2016tian} Tian, H., Xu, Z., He, J., \& Madsen, C.\ 2016, \apj, 824, 96
\bibitem[Uitenbroek(2001)]{2001uitenbroek} Uitenbroek, H.\ 2001, \apj, 557, 389 
\bibitem[Uitenbroek(2002)]{2002uitenbroek} Uitenbroek, H.\ 2002, \apj, 565, 1312 
\bibitem[Vernazza et al.(1981)]{1981vernazza} Vernazza, J.~E., Avrett, E.~H., \& Loeser, R.\ 1981, \apjs, 45, 635
\bibitem[Vissers et al.(2013)]{2013vissers} Vissers, G.~J.~M., Rouppe van der Voort, L.~H.~M., \& Rutten, R.~J.\ 2013, \apj, 774, 32
\bibitem[Vissers et al.(2015)]{2015vissers} Vissers, G.~J.~M., Rouppe van der Voort, L.~H.~M., Rutten, R.~J., Carlsson, M., \&
De Pontieu, B.\ 2015, \apj, 812, 11
\bibitem[Watanabe et al.(2008)]{2008watanabe} Watanabe, H., Kitai, R., Okamoto, K., et al.\ 2008, \apj, 684, 736
\bibitem[Watanabe et al.(2011)]{2011watanabe} Watanabe, H., Vissers, G., Kitai, R., Rouppe van der Voort, L., \& Rutten, R.~J.\ 2011, \apj, 736, 71
\bibitem[Xu et al.(2016)]{2016xu} Xu, Y., Cao, W., Ding, M., et al.\ 2016, \apj, 819, 89 
\bibitem[Yang et al.(2013)]{2013yang} Yang, H., Chae, J., Lim, E.-K., et al.\ 2013, \solphys, 288, 39
\bibitem[Yang et al.(2016)]{2016yang} Yang, H., Chae, J., Lim, E.-K., et al.\ 2016, \apj, 829, 100
\end{thebibliography}
\end{document}